\documentclass[runningheads]{llncs}
\usepackage[T1]{fontenc}
\usepackage{graphicx}
\usepackage{amsmath}
\usepackage{amssymb}
\usepackage{booktabs}

\usepackage{amsthm}
\usepackage{bm}
\usepackage{bbm}
\usepackage{colortbl}
\usepackage{array}
\usepackage{diagbox}
\usepackage{booktabs}
\usepackage{tikz}
\usepackage{fbox}
\usepackage{makecell}

\definecolor{cvprblue}{rgb}{0,1,0}
\usepackage[pagebackref,breaklinks,colorlinks,citecolor=cvprblue]{hyperref}
\usepackage[misc]{ifsym}

\usepackage{color}

\begin{document}
\title{Force Sensing Guided Artery-Vein Segmentation via Sequential Ultrasound Images}
\author{Yimeng Geng\inst{1,2} \and
Gaofeng Meng\inst{1,2,3} \and
Mingcong Chen\inst{3,4} \and
Guanglin Cao\inst{1,2,3} \and
Mingyang Zhao\inst{3}\and
Jianbo Zhao\inst{1,2} \and
Hongbin Liu\inst{1,3,5}$^{(\textrm{\Letter})}$
} 
\authorrunning{Y. Geng et al.}
% First names are abbreviated in the running head.
% If there are more than two authors, 'et al.' is used.
%
\institute{State Key Laboratory of Multimodal Artificial Intelligence Systems, Institute of Automation, Chinese Academy of Sciences, Beijing, China \and
School of Artificial Intelligence, University of Chinese Academy of Sciences \and
Centre for Artificial Intelligence and Robotics, HK Institute of Science \& Innovation, Chinese Academy of Sciences, Hong Kong SAR \and
City University of Hong Kong, Hong Kong SAR \and
School of Biomedical Engineering and Imaging Sciences, King’s College London, London, UK\\
\email{liuhongbin@ia.ac.cn}}
\maketitle              % typeset the header of the contribution
\begin{abstract}
Accurate identification of arteries and veins in ultrasound images is crucial for  vascular  examinations and interventions in  robotics-assisted surgeries. However, current methods for ultrasound vessel segmentation  face challenges in distinguishing between arteries and veins due to their morphological similarities. 
To address this challenge, this study introduces a novel force sensing guided segmentation approach to   enhance artery-vein segmentation accuracy by leveraging their distinct deformability. Our proposed method utilizes force magnitude to identify key frames with the most significant vascular deformation in a sequence of ultrasound images. These key frames are then integrated with the current frame through attention mechanisms, with weights assigned in accordance with force magnitude. Our proposed force sensing guided framework can be seamlessly integrated into various segmentation networks and achieves significant performance improvements in multiple U-shaped networks such as U-Net, Swin-unet and Transunet. Furthermore, we contribute the first multimodal ultrasound artery-vein segmentation dataset, Mus-V, which encompasses both force and image data simultaneously. The dataset comprises 3114 ultrasound images of carotid and femoral vessels extracted from 105 videos, with corresponding force data recorded by the force sensor mounted on the US probe. Our code and dataset will be publicly available.

\end{abstract}
\section{Introduction}
Vascular intervention surgery (VIS) is  a minimally invasive surgical technique involving catheter or device insertion into blood vessels \cite{vis1}.  This real-time and low-risk surgery plays a crucial role in treating various  diseases such as coronary artery diseases \cite{coronary} and 
peripheral vascular diseases \cite{peripheral}. Accurate segmentation of arteries and veins is vital in VIS, enabling clinicians to precisely identify target vessels for intervention and minimizing the risk of inadvertent injury to adjacent structures \cite{vis}. As a non-invasive, radiation-free and real-time modality, ultrasound imaging is widely used in VIS \cite{us-in-vis}.  Therefore, artery-vein segmentation in ultrasound images is crucial for enhancing the safety and efficacy of VIS. 

Ultrasound vessel segmentation, a subset of medical image segmentation, is developing rapidly, primarily because of the rapid evolution of deep learning techniques \cite{us-seg}.
In particular, a series of U-shaped networks have demonstrated outstanding performance \cite{unet,unet+++,dense-unet,swin-unet,transunet} and been widely applied in ultrasound vessel segmentation. Xie et al. \cite{vessel-seg-un} directly utilizes UNet \cite{unet} to segment carotid artery lumen.  Blanco et al. \cite{vessel-seg-ushape} presents a modified UNet to exploit adjacency information in spatially neighboring ultrasound images for  intravascular ultrasound images segmentation.  Groves et al. \cite{vessel-seg-av} 
 employs both Mask-RCNN \cite{mask-rcnn} and UNet to segment carotid artery (CA) and internal jugular vein (IJV).

% The encoder-decoder structure of UNet \cite{unet}, along with the skip connection, effectively utilizes image features at different scales. Transunet \cite{transunet} introduces a hybrid encoder based on the U-Net model, combining CNN and Transformer \cite{transformer} to address the limitations of traditional convolutional neural networks in modeling long-range dependencies and processing large-scale images. 
% Swin-unet \cite{swin-unet} is the first purely Transformer-based U-shaped architecture, utilizing Swin Transformer \cite{swin-transformer} blocks to extract hierarchical features from input images.

However, current ultrasound vessel segmentation networks either solely focus on morphological features within a single image \cite{vessel-seg-av}, making it challenging to differentiate arteries and veins that are morphologically similar, or only segment vessels without distinguishing between arteries and veins \cite{vessel-seg-fcn,vessel-seg-un,vessel-seg-ushape}. In clinical practice, doctors typically differentiate arteries from veins by palpating the patient's skin and observing the degree of vascular deformation in ultrasound images \cite{artery-vein}. During this process, arteries exhibit minimal deformation under compression, while veins undergo significant deformation and can even be completely occluded \cite{artery-vein}. Inspired by this method, we propose a novel force sensing guided approach for ultrasound artery-vein segmentation.This method aims to precisely segment arteries and veins using tactile force data, especially in scenarios involving robotic ultrasound systems.%examination or intervention. 

 Robotic Ultrasound Systems (RUS)  can improve the accuracy and safety of VIS through accurate image guidance and robotic arm operation, and provide better operating experience for doctors \cite{RUS,RUS1,robot-in-vis1,robot-in-vis2}. We set up a simple RUS in our experiments as shown in Fig. \ref{fig1}(a). The sensor mounted on the US probe can collected real-time force data, which is employed as a prior to identify key frames that represent maximum deformation.  These key frames effectively capture and store vascular deformation information, aiding in distinguishing the distinct deformability of arteries and veins as shown in Fig. \ref{fig1}(b). Then we can retrieve the stored information by computing the correlation between  key frames and the current frame to get enhanced features.  To validate the effectiveness of the proposed method, we also contribute a benchmark dataset containing both ultrasound images and force data called Mus-V. 

The main contributions are:
(1) We present a novel force sensing guided segmentation approach that utilizes force data to assist in artery-vein segmentation, leading to  enhanced segmentation accuracy. 
(2) Our method is versatile and demonstrates substantial performance improvement across multiple segmentation networks. 
(3) We contribute the first publicly available ultrasound artery-vein segmentation dataset with force data, providing a baseline for future researches.

\section{Ultrasound Vascular Dataset with Force data} 
\begin{figure}[t]
\begin{center}
  \includegraphics[width=\textwidth]{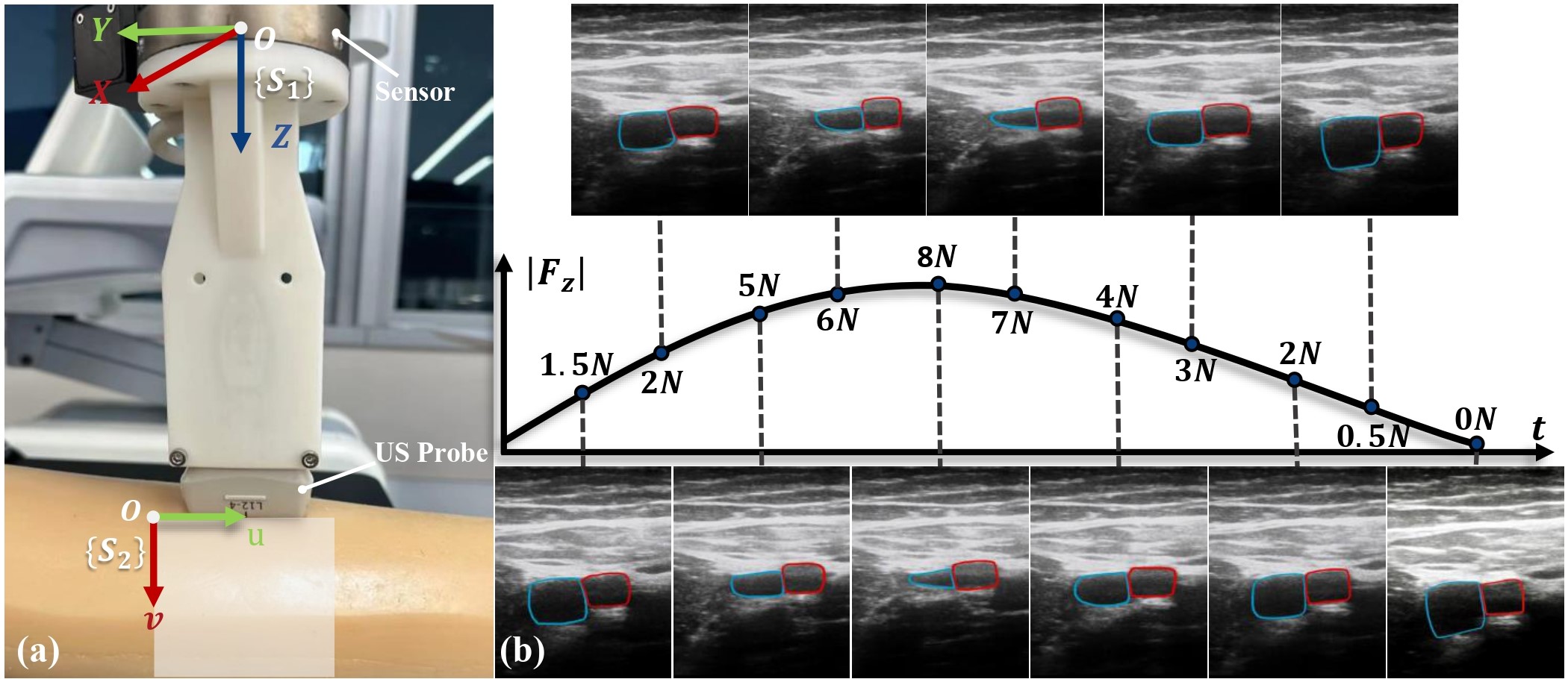}
\end{center}
\caption{(a) Hardware setup: a robotic arm, a force/torque sensor and an ultrasound probe. $\left \{  S_1\right \} $ and $\left \{  S_2\right \} $ respectively stand for  coordinate systems of the sensor and the US image  (b) An illustration of  data acquisition. The force along the z-axis, $|F_z|$, initially increases and then decreases over time $t$. Twelve points are selected along the curve, with corresponding force magnitudes and ultrasound scans. Veins are labelled in blue, while arteries are labelled in red. Notably, veins exhibit higher deformability compared to arteries.} \label{fig1}
\end{figure}

We introduce Mus-V: the first ultrasound artery-vein segmentation dataset containing both force and image data. The dataset collects ultrasound videos of carotid and femoral blood vessels, with the arteries and veins labelled separately for vascular analysis and identification. In addition, the dataset also records the force data collected by the force sensor mounted on the ultrasound probe during the collection of ultrasound videos. The structure of the robotic ultrasound system consists of an ultrasound (US) probe, a robotic arm carrying the US probe and a force/torque sensor mounted on the US probe (see Fig. \ref{fig1}(a)).  In each data collection session, an ultrasound video is recorded in conjunction with corresponding force data.  The contact force between the US probe and the skin is systematically adjusted to vary gradually. Meanwhile, the deformation of blood vessels is also strengthened or reduced as Fig. \ref{fig1}(b) shows. Subsequent to acquisition, the ultrasound videos and force data undergo aligning, cleaning, downsampling, and annotation.

A total of 3114 ultrasound images were sampled from 105 videos. The training and validation datasets comprise 2203 and 911 images, respectively. The force data is organized in a matrix format, with each row corresponds to an ultrasound image. Each column, from the first to the last, represents the force components in the X, Y, and Z directions, as well as the moment components in the X, Y, and Z directions, respectively.

\section{Method}
\subsection{Overview}
The most obvious difference between arteries and veins in US images lies in the extent of deformation when contact force varies. Therefore, capturing the deformation of vessels over time is crucial for accurate segmentation, which requires to compare different frames within the same video. Deformation magnitude correlates directly with force magnitude, thereby we utilize contact force data to identify key frames for reference. The selected key frames exhibit the most substantial degree of vascular deformation. Specific method for selecting key frames is described in Sec. \ref{sec.frame}. These key frames aid in  differentiation between arteries during the subsequent segmentation process.

\begin{figure}[t]
\includegraphics[width=\textwidth]{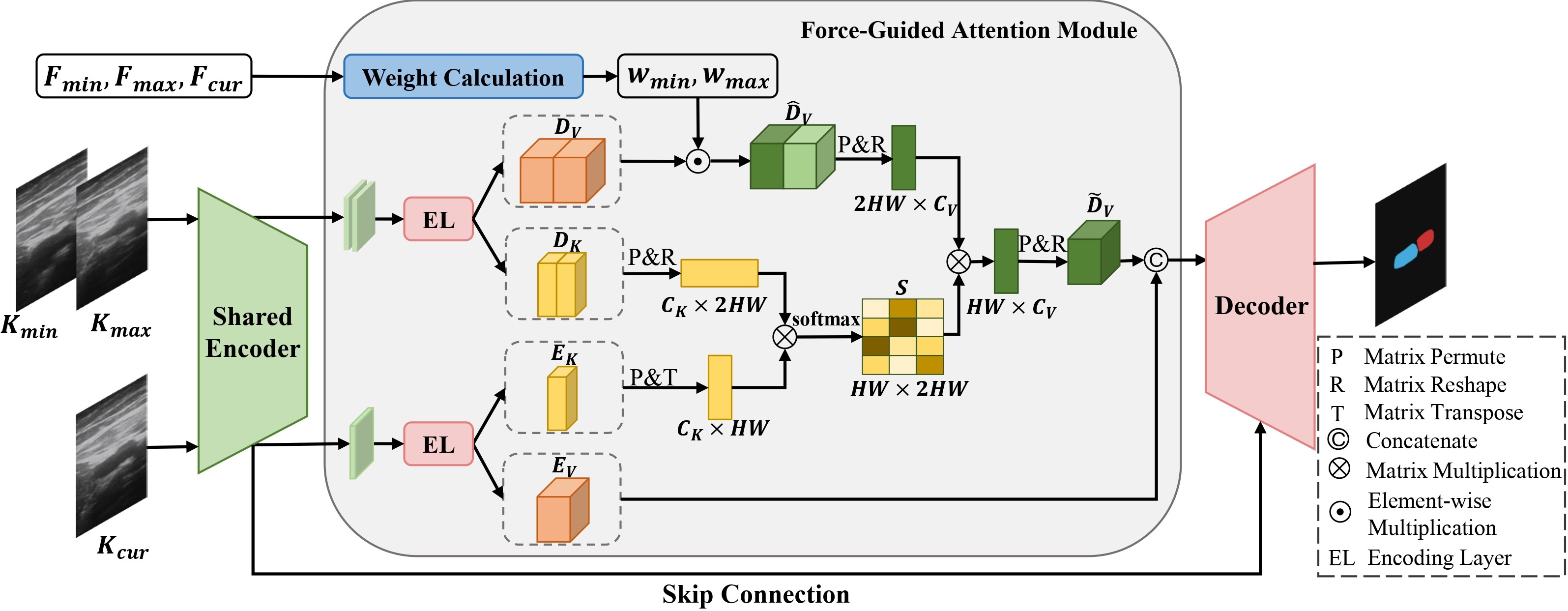}
\caption{Illustration of our proposed Force Sensing Guided Segmentation Framework. Selected key frames and the current frame are fed into a shared encoder to extract features. The force-guided attention module captures  relation between embeddings of key frames and current frames, generating an enhanced embedding $\tilde{D}_V$. The value embedding of current frame $E_V$ is concatenated with $\tilde{D}_V$ to get final segmentation result through the decoder. Skip connection is applied between the encoder and the decoder. } \label{fig2}
\end{figure}

Image segmentation networks based on the UNet architecture typically consist of an encoder that downsamples features and a decoder that upsamples features. Skip connections are utilized between different stages of the encoder and the decoder to fully leverage information at various resolutions. In the proposed fusion approach, key frames and the current segmentation frame are firstly fed into a shared encoder for feature extraction, with the final stage of the encoder outputting features with low resolution and high channel numbers. Subsequently, these features are processed by our proposed force-guided attention module. The fused features then enter the decoder to restore the original resolution and obtain the mask. This design allows for the encoder and decoder to be replaced with any U-shaped network, thereby enhancing the versatility of the method.
\subsection{Force-Driven Keyframe Selection}\label{sec.frame}
In the Robotic Ultrasound System, magnitude of contact force varies continuously, and changes between two consecutive frames may not be particularly evident. The magnitude of deformation is directly positively correlated with  magnitude of force along the z-axis. Therefore, we utilize force along the z-axis to select key frames that contain the most significant deformation information.  The target ultrasound image for segmentation is denoted as $K_{cur}$.  $K_{cur}$ is one frame from an ultrasound video with a duration of $T$ and $\mathcal{K}=\left \{ K_0,\dots,K_T \right \}$ indicates  the set of all the frames in the same video. Correspondingly,  $\mathcal{F}=\left \{ F_0,\dots,F_T \right \} $ represents the absolute value of force along the z-axis collected with $\mathcal{K}$.

As illustrated in Eq. \ref{eq1}, we select the image frame corresponding to the maximum force, denoted as $K_{\max}$, and the frame corresponding to the minimum compression force, denoted as $K_{\min}$. These frames, $K_{\min}$ and $K_{\max}$ are referred to as key frames.  

\begin{equation}\label{eq1}
    \left\{\begin{array}{ll} 
  K_{\min} = K_{\arg\min_{t}  F_t} \\
  K_{\max} = K_{\arg\max_{t}  F_t} 
\end{array}\right. 
\end{equation}
Therefore, $K_{\min}$ corresponds to the most dilated form of the blood vessel, while $K_{\max}$ corresponds to the most compressed form.
\subsection{Force-Guided Attention Module}
The output of the encoder has high-dimensional channels which are computationally expensive. Inspired by \cite{tmanet}, we firstly apply an encoding layer to the output for channel reduction. As illustrated in Fig. \ref{fig2}, the encoding layer including  a 1x1 convolution and a 3x3 convolution will generate a key feature and a value feature. We respectively stack the key features and value features related to key frames $K_{\min}$ and $K_{\max}$, and obtain a 4-dimension key feature $D_{V}\in R^{N\times C_K\times H \times W}$ and a 4-dimension value feature $D_K \in R^{N\times C_V\times H \times W}$, where $N=2$. $E_{K}$ and $E_{V}$ are the corresponding  key and value feature related to $K_{cur}$.

For frames in  proximity to a key frame, the reference value of the key frame is negligible. For example, if $K_{cur} = 
K_{\max}$, then $K_{\max}$ will fail to provide any meaningful information. Therefore, we propose a force-based dynamic weighting strategy, assigning a higher weight to the key frame with lower similarity to the current frame. Detailed calculation of weights is illustrated in  Sec. \ref{sec.weights}.

%%%%TODO
After obtaining force-based weights $w_{\min}$ and $w_{\max}$, we perform element-wise multiplication between the weights and $D_{V}$ to get weighted values $\hat{D}_{V}$. Then, $E_{K}$ and $E_{V}$  are transposed and reshaped to $E_{K} \in R^{C_K \times M}$ and $E_{V}\in R^{C_V \times M}$, $M=H\times W$. Meanwhile, $D_{K}$ and $\hat{D}_{V}$ are  permuted and reshaped to $\hat{D}_{K} \in R^{C_K \times NM}$ and $\hat{D}_{V}\in R^{C_V \times NM}$.

Subsequently, we conduct matrix multiplication between $D_{K}$ and $E_{K}$, and utilize a softmax layer to compute the attention map $S\in R^{N\times M}$,
\begin{equation}
    S_{ij}=\frac{\exp{(D^i_K\cdot E^j_K)}}{\sum_{j=1}^E\exp{(D^i_K\cdot E^j_K)}} 
\end{equation}
A larger value of $S_{ij}$ indicates that the $i^{th}$ position in $D_K$ has a greater influence on the $j^{th}$ position in $E_K$
Memory. After obtaining the attention map $S$, we multiply $S$ and $\hat{D}_{V}$ to get the force-guided enhanced feature $\tilde{D}_V$.
Finally, we combine $\tilde{D}_V$ with the value feature of the current frame. For simplicity, we utilize feature concatenation to get the aggregated multimodal feature $f$ as follows:
\begin{equation}
    f = concat(\tilde{D}_V,E_V)
\end{equation}

\subsection{Force-Based Dynamic Weights}\label{sec.weights}
Adhering the principle that a higher weight is assigned to the key frame with lower similarity to the current frame, we design a weight calculation formula  based on  force magnitudes. The dynamic weights are directly proportional to the difference in force magnitudes and lies within the range of $\left [ 0,1 \right ] $,
\begin{equation}
\label{w1}
    w_{\min}=\frac{F_{cur}-F_{\min}}{F_{\max}-F_{\min}}
\end{equation}
\begin{equation}
\label{w2}
    w_{\max}=\frac{F_{\max}-F_{cur}}{F_{\max}-F_{\min}}
\end{equation}
where  $w_{\min}$ and $w_{\max}$ are  force-based dynamic weights for $K_{\min}$ and $K_{\max}$. For each input, a new pair of force-based weights is computed and applied. $F_{cur}$, $F_{\min}$, $F_{\max}$ are force respectively corresponding to $K_{cur}$, $K_{\min}$ and $K_{\max}$.
\section{Experiments}
\subsection{Implementation and Setup  }
Our experiments are implemented based on PyTorch on a RTX 4090 GPU.  We employ the ReduceLROnPlateau \cite{reduce} 
policy to automatically adjust the learning rate and utilize RMSprop \cite{rmsprop} as the optimizer. Momentum, weight decay and batch size are set to 0.9, 1e-8 and 4 for all experiments, respectively. 
We set the maximum number of epochs as 120 for experiments. For data augmentation, we apply random horizontal and vertical flip, random intensity transformation, random translation, random rotation  for input images and key frames for all experiments. All images are resized to 256x256.
We also utilize Torchio \cite{torchio} to add random bias fields and random noise to input images and key frames. To address the issue of class imbalance, we employed weighted cross-entropy loss with weights of 1, 30.7, and 23.1 for background, arteries and veins, respectively. 

\subsection{Comparison with State-of-the-Arts}
\begin{figure}[t]
\includegraphics[width=\textwidth]{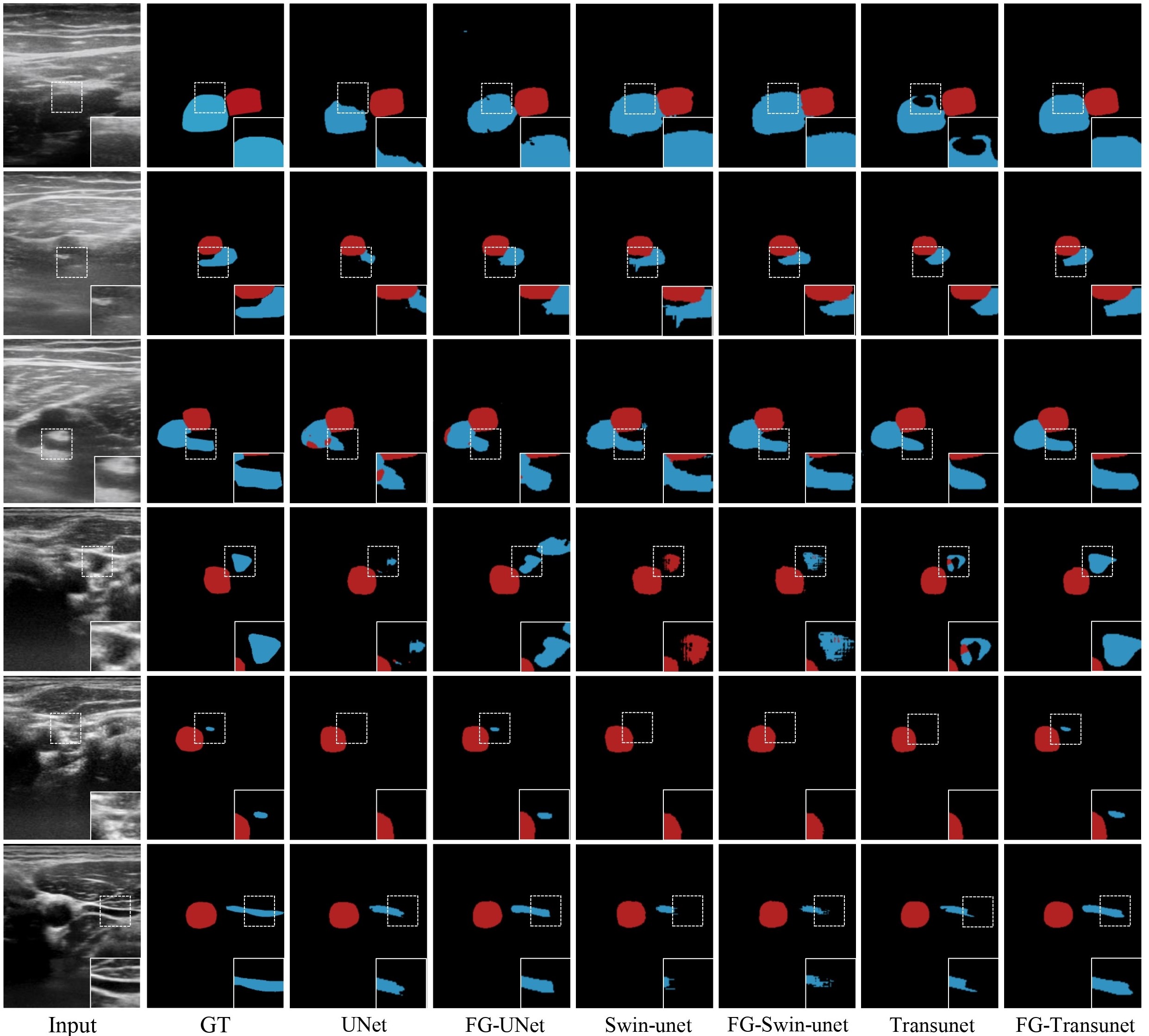}
\caption{Qualitative results. The results demonstrate that  integrating force sensing guidance effectively enhances the performance of three baseline networks. Specifically, our method improves the model's ability to differentiate between arteries (red) and veins (blue), and ability to segment small veins.} \label{fig3}
\end{figure}

We conduct comprehensive experiments by applying force sensing guidance to three different U-shaped networks, namely UNet, Swin-unet and Transunet. The force-guided attention module is inserted as a ``neck'' between the encoder and the decoder of three basic networks. For Transunet, we choose to use the R50-Vit-B16 version. Table \ref{tab1} displays the performance metrics and FLOPs of all the methods. The network applied with force sensing guidance is named with the prefix ``FG'' followed by the network name.  In conclusion, Force Sensing Guided TranUnet (FG-Transunet) achieves the highest performance, attaining  75.63\% mIoU and 0.8547 dice coefficient. Additionally, the application of force sensing guidance results in performance enhancements of 1.2\%, 2.79\% and 1.25\% mIoU in Swin-UNet, UNet and Transunet, respectively . Qualitative results are presented in Fig. \ref{fig3}
\begin{table}[t]
\caption{Comparison results of different methods on the Mus-V dataset.} \label{tab1}
\begin{center}
\begin{tabular}{p{0.25\textwidth}|p{0.15\textwidth}|p{0.22\textwidth}|p{0.18\textwidth}}
\Xhline{0.7pt}
Method & mIoU(\%)$\uparrow$ & Dice Coefficient$\uparrow$ & FLOPs(M)$\downarrow$\\
\Xhline{0.5pt}
Swin-unet\cite{swin-unet} &  68.36 & 0.7968& 0.774$\times 10^4$\\
UNet\cite{unet} &    69.23  & 0.8140& 4.019 $\times 10^4$\\
Transunet\cite{transunet} & 74.38 & 0.8499& 3.225$\times 10^4$\\
\Xhline{0.5pt}
FG-Swin-unet & 69.56 & 0.8033& 1.855$\times 10^4$\\
FG-UNet &    72.02  & 0.8280& 8.016$\times 10^4$\\
FG-Transunet  & \textbf{75.63} & \textbf{0.8547}& 8.867$\times 10^4$\\
\Xhline{0.7pt}
\end{tabular}
\end{center}
\end{table}  
\subsection{Ablation Studies}
We conduct several ablation experiments to validate the effectiveness of all the components of the proposed approach, utilizing UNet \cite{unet} as our baseline. To save computational resources and training time, we also exploit Force Sensing Guided UNet (FG-UNet) as the base model for all ablation experiments.
\begin{table}[t]
\caption{Ablation study results. The w/o KFS represents that force-driven key frame selection is removed and two frames temporally before the current frame are selected as key frames. The w/o FBW means that force-based dynamic weights are removed. }\label{tab2}
\begin{center}
\begin{tabular}{p{0.35\textwidth}|p{0.15\textwidth}|p{0.22\textwidth}}
\Xhline{0.7pt}
Method & mIoU(\%)$\uparrow$ & Dice Coefficient$\uparrow$\\
\Xhline{0.5pt}
UNet\cite{unet} &    69.23  & 0.8140\\
FG-UNet  w/o KFS\&FBW &   70.22  & 0.8205\\
% tau-1 &  71.48 & 0.1763 \\
FG-UNet  w/o FBW& 71.36 & 0.8244\\
FG-UNet & \textbf{72.02} & \textbf{0.8280}\\
\Xhline{0.7pt}
\end{tabular}
\end{center}
\end{table}

%\subsubsection{Force-based Weights}
To validate the effectiveness of the  force-based dynamic weights, we conducted experiments by removing the weights multiplication. The experimental results are shown in Table \ref{tab2}. After removing the force-based weights, the mIoU decreases by 0.66\%.

%\subsubsection{Key Frame Selection}
To investigate the effect of the proposed force-driven key frame selection strategy, we conduct an experiment that simply selecting two frames preceding the current frame in time as key frames. The results are displayed in Table \ref{tab2}. The performance has decreased significantly in the aspect of both mIoU and dice coefficient  after removing the key frame selection. The mIoU decreases by 1.14\% and  the dice coefficient  decreases by 0.039.
\section{Conclusion}
In this study, we introduced a pioneering force sensing guided approach for ultrasound artery-vein segmentation. Our method includes innovative components such as force-driven key frames selection, force-guided attention module and force-based dynamic weights. Extensive experiments demonstrated that our proposed approach led to   notable performance enhancements across multiple U-shaped networks. Furthermore, we presented Mus-V, the first publicly available ultrasound artery-vein  segmentation dataset including images and force. Future directions include further optimizing the performance by exploring additional methodologies, as well as expanding the capacity of Mus-V.

%
% ---- Bibliography ----
%
% BibTeX users should specify bibliography style 'splncs04'.
% References will then be sorted and formatted in the correct style.
%
\bibliographystyle{splncs04}
\bibliography{ref}

\end{document}